\documentclass[reprint, sort&compress]{elsarticle}

\usepackage{lineno,hyperref}

\usepackage{graphicx}
\usepackage{dcolumn}
\usepackage{bm}
\usepackage{upgreek}
\usepackage{hyperref}
\usepackage{ams math}
\usepackage[margin=1.2in]{geometry}

\newcommand{\Nu}{N\!u}%
\journal{International Journal of Heat and Mass Transfer}

\begin{document}

\begin{frontmatter}

\title{Heat transfer enhancement by mist/air two-phase flow in a high-temperature channel}


\author[TJU]{Junxian Cao}
\author[TJU]{Mengqi Ye}
\author[BH]{Haiwang Li}
\author[TJU]{Tianyou Wang}
\author[TJU]{Zhizhao Che\corref{cor1}}
\cortext[cor1]{Corresponding author.
}
\ead{chezhizhao@tju.edu.cn}
\address[TJU]{State Key Laboratory of Engines, Tianjin University, Tianjin, 300072, China.}
\address[BH]{National Key Laboratory of Science and Technology on Aero Engines Aero-Thermodynamics, Beihang University, Beijing 100191, China}


%
%

\begin{abstract}
Mist/air two-phase flow is a promising cooling technique for many applications such as internal cooling of gas turbine blades. A significant enhancement of heat transfer can be achieved with a low mass fraction of droplets by utilizing the latent heat of the droplets. Using newly designed atomizers to accurately control the mist droplets, this study experimentally explores the heat transfer performance of mist/air flow in a high-temperature channel with a maximum temperature of 880 K. The effects of the mist/air mass ratio, droplet size, Reynolds number, and wall heat flux are studied. The results show that the cooling performance of the test section can be significantly improved by even adding a small amount of droplets. Considering mist droplets of different sizes, larger droplets can cause more remarkable temperature reduction, while smaller droplets can improve the uniformity of temperature distribution. For large droplets, the cooling effect in the upstream is more obvious than that in the downstream due to the interaction between the wall and the droplets, and with the increase of mist/air mass ratio, the area with obvious cooling extends downstream. The performance of mist/air cooling is tested by increasing the heat flux until the maximum temperature at the outlet reaches a predetermined value. Compared with air-only cooling, the increment in the wall heat flux by the mist/air cooling with a mass ratio of 3\% can be up to 18.4\%.
\end{abstract}

\begin{keyword}
\texttt
{Mist/air cooling \sep High-temperature channel \sep Heat transfer enhancement \sep Two-phase flow}
\end{keyword}

\end{frontmatter}


\def \scaleSize {0.8}
\def \scaleSiz2 {0.6}

\section{Introduction}\label{sec:sec1}
The effective cooling of high-temperature channels is important in many applications, such as the cooling of turbine blades in gas turbines and the cooling of reactor cores in nuclear reactors. For example, in gas turbines, increasing the turbine inlet temperature can effectively improve its output power \cite{Han2012CoolingTechnology}, but the high temperature poses a challenge to turbine blades. Currently, the inlet temperatures of advanced gas turbines are higher than 1700$^\circ$C, which is much higher than the yield point temperature of the blade material of 1200$^\circ$C \cite{Han2018AwardPaper}, hence it is extremely important to cool turbine blades. The cooling methods of turbine blades include impact cooling, film cooling, transpiration cooling, internal convective cooling, etc. To further improve the turbine efficiency, it is of great significance to develop efficient cooling technology. As an effective cooling method, internal cooling of the blades has a huge potential and application demand \cite{Han2013HeatTransfer}.

To improve the cooling performance for internal blade cooling, several internal cooling strategies have been proposed, such as air-only cooling, closed-steam cooling, two-phase fluid mist/steam, and mist/air cooling \cite{Zeng2016DifferentRibAngles}. The benefits from air-only cooling method are limited due to the poor thermal conductivity \cite{Shi2013SteamAsCoolant}. Because the specific heat of water steam is twice as much as that of air, closed-steam cooling can effectively improve the cooling effect of gas turbine blades \cite{Corman1996DevelopmentStatus}, but steam-cooling usually requires a large amount of steam (3\%-5\%) \cite{Guo2000ExperimentalSystem} to meet the cooling requirements, which directly reduces the thermal efficiency of turbines. Two-phase flow cooling, in which small droplet particles are injected into the steam or air, has many advantages \cite{Wang2013TurbineAirfoils}. It not only has larger specific heat, but also can utilize the latent heat of vaporization. In addition, the droplets can disturb the flow, particularly at the boundary layer, hence enhancing the mass, momentum, and energy transfer.

Because of the great cooling potential, mist/steam flows have been studied in some experiments. Guo et al.\ \cite{Guo2000TubeBend, Guo2000ResultsAndModeling} studied the mist/steam flow and heat transfer in a heated horizontal tube and a 180-degree tube bend. The results showed that the addition of droplets to the main flow significantly improved the heat transfer of steam. The average increase in the heat transfer capacity of the straight tube was 100\% at a mist/steam mass ratio below 5\%. In the bent tube, the outer surface of the bend was cooled more effectively than the inner surface. Li et al.\ \cite{Li2001JetImpingement, Li2001ImpingementJet, Li2003ImpingementJets, Li2003ConcaveSurface} studied mist/steam impingement cooling, and found that the heat transfer could be enhanced by up to 200\% in the stagnation region by injecting only 1.5\% mist by mass. Shi et al.\ \cite{Shi2019SmoothChannel} found that the heat transfer capacity in the upstream of a channel was much greater than that in the downstream due to the accumulation of droplets at the entrance region of the test section.

Some numerical simulations of the heat transfer in mist/steam flow have also been performed. The results of Jiang et al.\ \cite{Jiang2020ColumnRowRibs} showed that the effects of Reynolds number and the mist/steam mass ratio on the wall heat transfer coefficient were greater than those of the wall heat flux and the droplet diameter. The results of Zhang et al.\ \cite{Zhang2015RectangularChannel} showed that a large Reynolds number corresponds to a more pronounced heat transfer enhancement of mist/steam cooling. The study of Jiang et al.\ \cite{Jiang2019CoolingPassage} for a U-shaped channel revealed that the heat transfer effect was higher when the mist diameter was 10 $\upmu$m than that of 5 $\upmu$m and 15 $\upmu$m.

Compared with mist/steam cooling, the system of mist/air cooling can be much simpler because the steam generator is not required. It is particularly important for applications when the generation of steam is difficult. In addition, the evaporation of droplets in mist/air cooling is not affected by the total pressure of the main flow \cite{Zhao2014HeatTransfer}, hence the droplets in mist/air cooling are more likely to evaporate under the same temperature and the same pressure. Several studies of mist/air cooling have been reported. Mori et al.\ studied the mist/air cooling in a 1.8-mm-diameter channel \cite{Mori1982MistCooling}, and divided the wall temperature distribution into three regions, a region of slow temperature rise, followed by a region of sharp temperature rise, and a region with a temperature gradient equal to that for airflow. Zhao et al.\ \cite{Zhao2014DropletDynamics, Zhao2014HeatTransfer} used a phase Doppler particle analyzer (PDPA) to measure the droplet size and spatial distribution in film cooling. Their results showed the mist/air film cooling made the cooling coverage increase by 83\% compared to air-only film cooling, and the surface temperature was more uniform. To demonstrate the effectiveness of downstream mist/air cooling of the channel, Ragab et al.\ \cite{Ragab2018DropletDynamics, Ragab2018HeatTransfer} studied the effect of downstream cooling of a long channel (100 times of the diameter), and the results showed that, for mist/air cooling, the downstream cooling was still effective and mist/air cooling had a longer cooling length than air-only cooling.

It is worth noting that the existing experiments were conducted at temperatures ($<500$ K) far below the operating temperature of the gas turbine. The mist/air cooling under these relatively low temperature conditions is close to spray cooling because droplets tend to form a liquid film on the wall \cite{Zhao2014HeatTransfer}. There are only a few simulation studies are available at high temperatures. Dhanasekaran et al.\ \cite{Dhanasekaran2008HorizontalTube, Dhanasekaran2012BendTube, Dhanasekaran2013AngledRibTurbulators} simulated the actual gas turbine working condition, and the results suggested that the average cooling enhancement in the elevated operating condition was higher than that in the corresponding low pressure and temperature operating condition. The numerical simulation by Gao et al.\ \cite{Gao2017OperationalConditions} indicated that the heat transfer enhancement effect of mist/steam flow was reduced at high-temperature conditions.

It should also be noted that the generation and transport of mist are very important in mist/air cooling. Some researchers \cite{Sefko2015DispersedFlow} placed atomization nozzles in the transition section, where the droplets mixed with air and then entered the test section. The sidewall of the transition section had an outlet for extracting the deposited liquid. Some researchers \cite{Guo2000ExperimentalSystem, Wang2013TurbineAirfoils} utilized a mixing chamber to mix mist with air and then connected the mixing chamber to the test section through a transition section. The transition section was heated \cite{Guo2000ExperimentalSystem} or the large droplets were filtered \cite{Wang2013TurbineAirfoils} to reduce droplet deposition as they pass through the transition section. In those atomization methods, the mist/air mass ratio was calculated by collecting and weighing droplets that failed to enter the channel, hence it was difficult to achieve accurate control of the mist/air mass ratio.

In this paper, the mist/air cooling in a high-temperature channel is studied experimentally up to 880 K. New atomizers are designed to generate and transport mist/air two-phase flow, through which accurate control of the mist/air mass ratio is achieved. A series of experiments are conducted to investigate the effects of the droplet particle size, the mist/air mass ratio, the Reynolds number, and the wall heat flux. The comparison between mist/air cooling and air-only cooling is made. The capability of the mist/air cooling is also studied by fixing the maximum temperature and increasing the heat flux until the maximum temperature is reached. The performance of the mist/air cooling is quantitatively compared using the thermal performance ratio and the heat transfer non-uniformity.

\section{Experimental details}\label{sec:sec2}
The experimental system mainly consists of a heat transfer test system and a water atomization system. The heat transfer test system was designed to withstand high-temperature with controllable wall heat flux to study the heat transfer of mist/air two-phase flow under different wall heat fluxes. The atomization system was used to generate mist/air mixture with stable droplet distribution, controllable droplet sizes, and controllable flow rate.

\subsection{Heat transfer test system}\label{sec:sec21}
\begin{figure}
  \centering
  \includegraphics[width=0.55\columnwidth]{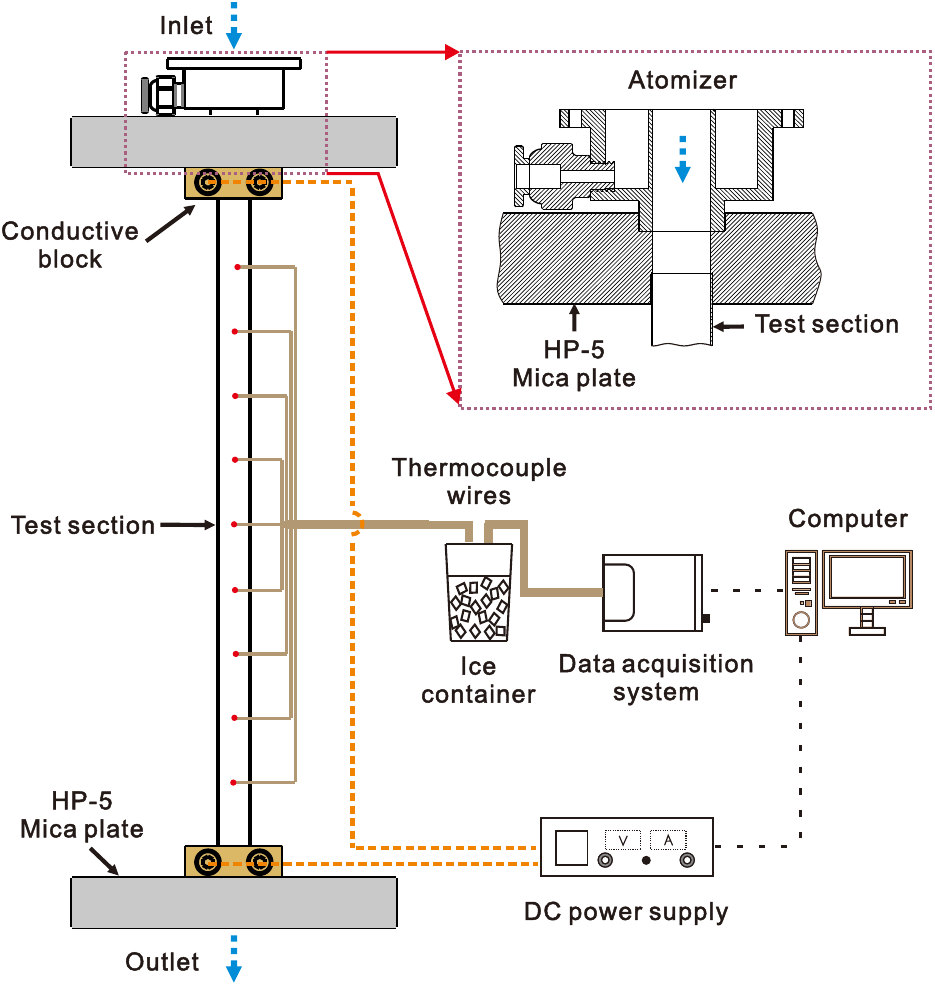}\\
  \caption{Schematic diagram of the heat transfer test system.}\label{fig:fig01}
\end{figure}

The heat transfer test system includes a heated test section and a data acquisition system, as shown in Fig.\ \ref{fig:fig01}. The heated test section is a straight tube made of 316L stainless steel with 10 mm in outer diameter, 230 mm in length, and 0.3 mm in wall thickness. The wall of the stainless-steel tube served as a resistance heater by connecting to a power supply. The heating part of the tube is 200-mm long, and on each end of the test section, there is a conductive block of 10 mm for the connection with the power supply and a short section of 5-mm inserted into a 15-mm-thick HP-5 mica plate. The main role of the mica plate is to connect the test section and the atomization system with good electrical and thermal insulation, and it can withstand temperature above 1000 K. The wall of the test section was connected to a programmable high-power DC power supply (ITECH IT6582D) with two 316L stainless steel conductive blocks. The wall heat flux of the test section was regulated by adjusting the power output of the DC power supply with an uncertainty of less than 0.02\%. Over the tube surface, a thin layer of 0.1-mm-thick mica paper was firmly wrapped with high-temperature cement (Omega CC). The temperature drop across the mica paper was neglected \cite{Guo2000ExperimentalSystem}. The exterior of the test section was covered by aluminum silicate insulating cotton of a thickness of 50 mm for thermal insulation.

Nine K-type SMD thermocouples (with 20-mm apart fixed with Omega CC high-temperature cement) were used to measure the temperature variation along the test section, as shown in Fig.\ \ref{fig:fig01}. Another thermocouple was placed on the centerline at the exit of the test section to measure the bulk fluid temperature. Ice-water mixture was used to provide a zero-degree environment for the cold-junction of thermocouples, and the extension wires from the cold-junction were further connected to a data acquisition card (NI USB-6361), through which the temperature readings were logged by a computer. Data were collected after the steady state was established as indicated by the reading of the thermocouples.

\subsection{Water atomization system}\label{sec:sec22}
\begin{figure}
  \centering
  \includegraphics[width=0.55\columnwidth]{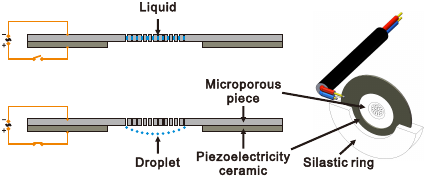}\\
  \caption{Schematic diagram of the microporous atomization unit.}\label{fig:fig02}
\end{figure}

The water atomization system mainly consisted of an air supply system, a water supply system, and microporous atomization units. For the air supply system, air was supplied at constant gauge pressure (3 bar) from a reciprocating piston compressor with a pressure reducing valve, and an airflow controller (ALICAT KM6001-21-1-00-1-500) was used to control the flow rate. In the water flow system, distilled water was supplied with a peristaltic pump (Longer L00-1S-2+YZII15) to control the water flow. The microporous atomization unit was mainly composed of a high-frequency drive circuit, a piezoelectric ceramic film, and a microporous stainless-steel sheet, as shown in Fig.\ \ref{fig:fig02}. The high-frequency AC circuit drove the piezoelectric ceramic to vibrate, which drove the microporous sheet to produce high-frequency vibration and atomize the liquid at the microporous sheet into a stream of droplets. In this study, the AC input voltage was 5 V and the frequency was 108 kHz. To adjust the size of the droplets, two types of microporous sheets were used: one with 2000 pores of 5 $\upmu$m in diameter, and the other with 1000 pores of 12 $\upmu$m in diameter.

\begin{figure}
  \centering
  \includegraphics[width=\columnwidth]{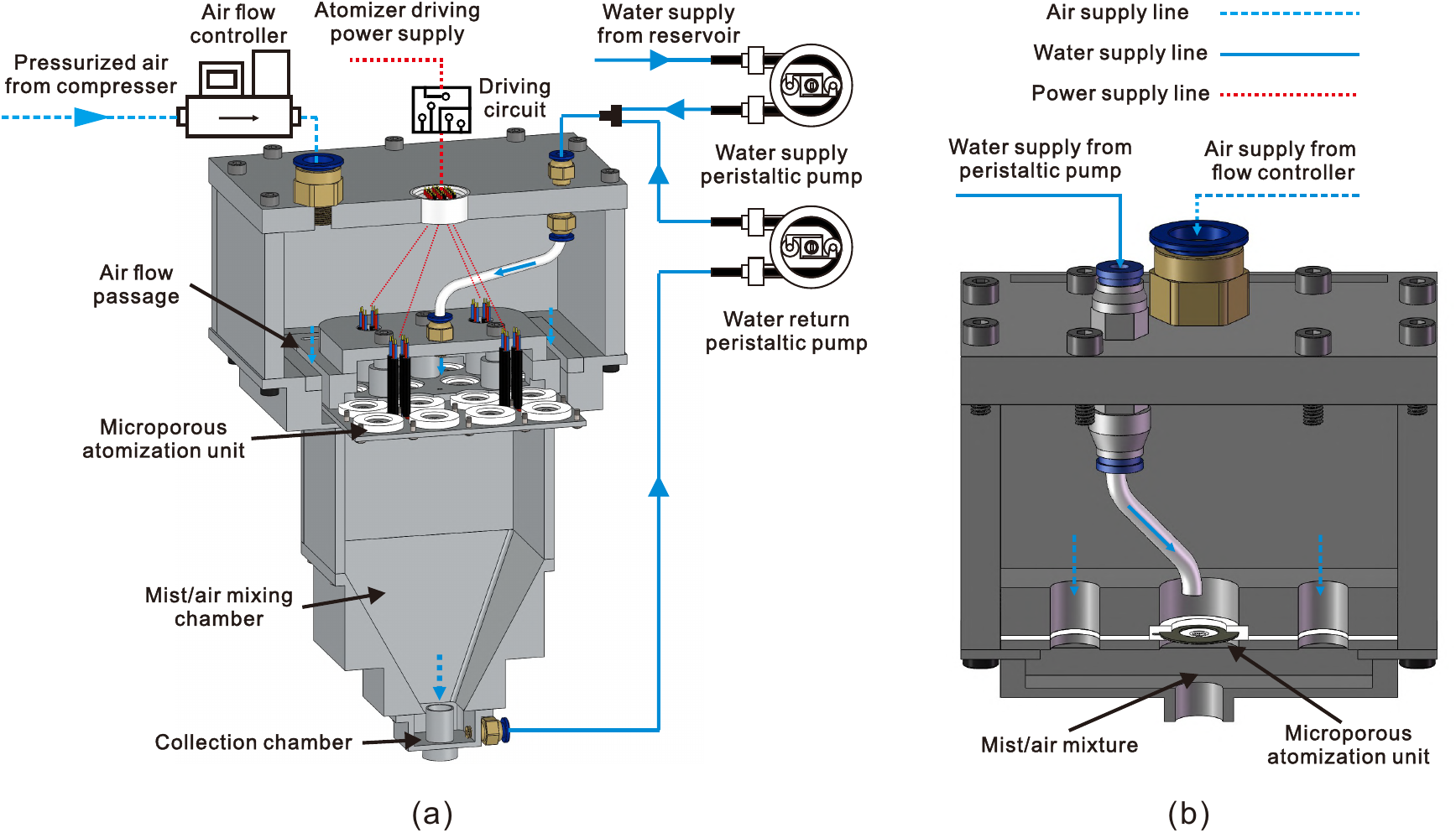}\\
  \caption{Schematic diagram of the two atomizers used in the experiment. (a) 5-$\upmu$m atomizer with 16 5-$\upmu$m atomization units; (b) 12-$\upmu$m atomizer with one 12-$\upmu$m atomization unit.}\label{fig:fig03}
\end{figure}
Considering the limitation in the atomization rate for one atomization unit, two atomizers were designed in this study: one atomizer had 16 5-$\upmu$m atomization units, and the other had only one 12-$\upmu$m atomization unit, as shown in Figs.\ \ref{fig:fig03}(a) and \ref{fig:fig03}(b), respectively. For the 5-$\upmu$m atomizer, water was supplied to the atomizer via the peristaltic pump with a controlled volume flow rate, and air was supplied from the airflow controller into the mixing chamber to mix thoroughly with the droplets produced by the microporous atomization units. Finally, the well-mixed mist/air flowed out of the mixing chamber and was supplied to the test section. During this process, some droplets were deposited on the walls of the mixing chamber and eventually flowed into a collection chamber. The water in the collection chamber was pumped back to the atomization units by another peristaltic pump. This return loop ensured that the droplet mass flow rate at the outlet of the mixing chamber was equal to the mass flow rate supplied from the water supply peristaltic pump, achieving accurate control of the atomization mass flow rate. For the 12-$\upmu$m atomizer, it had only one atomization unit and the droplet stream was well aligned with the test section. Therefore, the droplet stream could directly enter the test section without impingement on the mixing chamber, and the return loop was not needed for the 12-$\upmu$m atomizer.

\begin{figure}
  \centering
  \includegraphics[width=0.7\columnwidth]{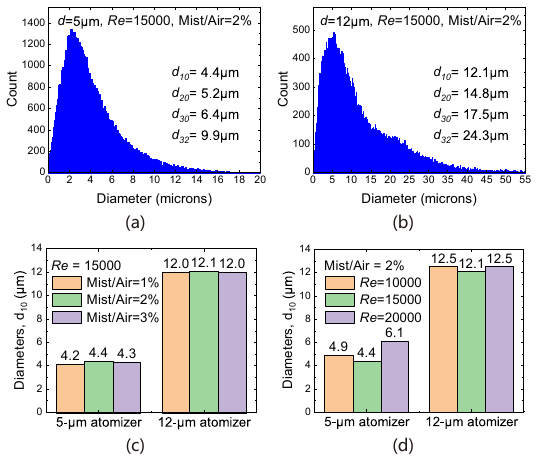}\\
  \caption{ Distribution and statistics of mist droplet size under different conditions. (a) Histogram of droplet size for the 5-$\upmu$m atomizer; (b) histogram of droplet size for the 12-$\upmu$m atomizer; (c) arithmetic means droplet diameter ($d_{10}$) for different mist/air ratios at a fixed Reynolds number of $Re = 15000$; (d) arithmetic means droplet diameter ($d_{10}$) for different Reynolds numbers at a fixed mix/air ratio of 2\%. }\label{fig:fig04}
\end{figure}

The droplet size at the outlet of the atomizers was measured using a phase Doppler particle analyzer (PDPA, Dantec Dynamics). The measurement was taken at a point on the axis of the atomizer and 10 mm below the atomizer outlet. The histograms of the droplet sizes for the two atomizers are shown in Figs.\ \ref{fig:fig04}(a) and \ref{fig:fig04}(b), respectively. Most of the droplets of the 5-$\upmu$m atomizer are smaller than 10 $\upmu$m; in contrast for the 12-$\upmu$m atomizer, about half of the droplets are within this range and the other half are between 10 $\upmu$m and 55 $\upmu$m. The arithmetic means droplet diameters ($d_{10}$) and the Sauter mean diameter ($d_{32}$) are also calculated for quantitative comparison of the droplet size. Generally, $d_{10}$ represents the diameters of most droplets, and $d_{32}$ represents the diameters of large droplets \cite{Guo2000ExperimentalSystem}. At the Reynolds number of 15000 and mist/air mass ratio of 2\%, the arithmetic means droplet diameters ($d_{10}$) is 4.4 $\upmu$m for the 5-$\upmu$m atomizer, and 12.1 $\upmu$m for the 12-$\upmu$m atomizer; the Sauter mean diameter ($d_{32}$) is 9.9 $\upmu$m for the 5-$\upmu$m atomizer, and 24.3 $\upmu$m for the 12-$\upmu$m atomizer. The arithmetic means droplet diameter ($d_{10}$) for different mist/air ratios and for different Reynolds numbers are also compared in Figs.\ \ref{fig:fig04}(c) and \ref{fig:fig04}(d). The results show that the droplet size varies only slightly as the Reynolds number and the mist/air mass ratio change. Therefore, the effects of the Reynolds number and the mist/air mass ratio on the droplet size are negligible.

\subsection{Heat transfer analysis}\label{sec:sec23}
The heat transfer coefficient and the Nusselt number are used to quantify the heat transfer of the mist/air flow on the inner surface of the channel. The Nusselt number is defined as
\begin{equation}\label{eq:eq01}
  \Nu=hD/\lambda,
\end{equation}
where $D$ is the inner diameter of the test section, and $\lambda $ is the thermal conductivity of dry air at the inlet of the test section (300 K). The heat transfer coefficient $h$ is calculated as follows
\begin{equation}\label{eq:eq02}
  h={{q}_{n}}/({{T}_{w,z}}-{{T}_{b,z}}),
\end{equation}
where ${{T}_{w,z}}$ and ${{T}_{b,z}}$ are the wall temperature and the bulk temperature at the axial location of $z$, respectively. ${{T}_{b,z}}$ is obtained by linear interpolation from the outlet and inlet temperatures, with the inlet temperature of 300 K and the outlet temperature measured experimentally. ${{q}_{n}}$ in Eq.\ (\ref{eq:eq02}) is the net wall heat flux, which is equal to the difference between the wall input heat flux $q$ and the heat loss flux at the wall ${{q}_{loss}}$. The wall input heat flux is calculated by the input power of the DC power supply as $q=UI$. Hence the net wall heat flux is
\begin{equation}\label{eq:eq03}
  {{q}_{n}}=q-{{q}_{loss}}=UI-{{q}_{loss}}.
\end{equation}

The heat loss flux at the wall ${{q}_{loss}}$ mainly consists of the radiative heat loss at the surface, the conductive heat loss via the insulation layer, and the conductive heat loss via the conductive blocks. In this study, the outside of the channel is wrapped by aluminosilicate insulation wool, which is a good thermal insulation material. The radiative heat loss and the conductive heat loss via the conductive blocks could be neglected because the insulation layer reduces the surface temperature and the conductive blocks are close to the inlet and outlet of the test section \cite{Guo2000ExperimentalSystem}. Therefore, the conductive heat loss via the insulation layer is the main heat loss, and has to be considered. To determine the heat loss via the insulation layer, a heat transfer experiment was performed by blocking the inlet and the outlet of the test section and measuring the temperature distribution without airflow. In the steady state, the heat generated by the electric heating is all lost by heat conduction via the insulation layer. Then a fitting relationship between the wall temperature and the heat loss can be obtained, i.e., ${{q}_{loss,z}}=a{{T}_{w,z}}+b$ with $a=12.5\text{W/}\left( {{\text{m}}^{\text{2}}}\cdot \text{K} \right)$, and $b=-4047.4\text{W/}{{\text{m}}^{\text{2}}}$. This fitting relationship was then used to calculate the local heat loss in the experiment.

The Reynolds number is adjusted in this study by controlling the volume flow rate of air
\begin{equation}\label{eq:eq04}
  Re=\rho uD/\mu,
\end{equation}
where $u$ is the velocity of air controlled by airflow controller and calculated from the flow rate. In Eq.\ (\ref{eq:eq04}), the physical parameters of dry air at the channel inlet temperature of 300 K is used to calculate the Reynolds number.

The heat transfer non-uniformity is used to quantify the heat transfer \cite{Jiang2020ColumnRowRibs} and it is defined as the normalized standard deviation of the heat transfer coefficient,
\begin{equation}\label{eq:eq05}
  \eta =\frac{1}{\overline{h}}\sqrt{\frac{1}{N}\sum\limits_{i=1}^{N}{{{({{h}_{i}}-\overline{h})}^{2}}}},
\end{equation}
where $\overline{h}$ is the average heat transfer coefficient of the test section and it is defined as
\begin{equation}\label{eq:eq06}
  \overline{h}=\frac{1}{N}\sum\limits_{i=1}^{N}{{{h}_{i}}}.
\end{equation}
A larger value of $\eta $ implies an increased non-uniformity of the heat transfer coefficient along the channel, which is the result of the combined effect of wall temperature variation and coolant temperature variation.

The thermal performance ratio ${{\Nu}_{m}}/{{\Nu}_{a}}$ is used to quantify the degree of heat transfer enhancement, where ${{\Nu}_{m}}$ and ${{Nu}_{a}}$ are the average Nusselt numbers of mist/air cooling and air-only cooling, respectively. The average Nusselt numbers are calculated from the average heat transfer coefficients in Eq.\ (\ref{eq:eq06}) using Eq.\ (\ref{eq:eq01}).

\section{Results and discussion}\label{sec:sec3}
\subsection{Comparison between mist/air cooling and air-only cooling and the effects of mist/air mass ratio and droplet diameter}\label{sec:sec31}
\begin{figure}
  \centering
  \includegraphics[width=0.4\columnwidth]{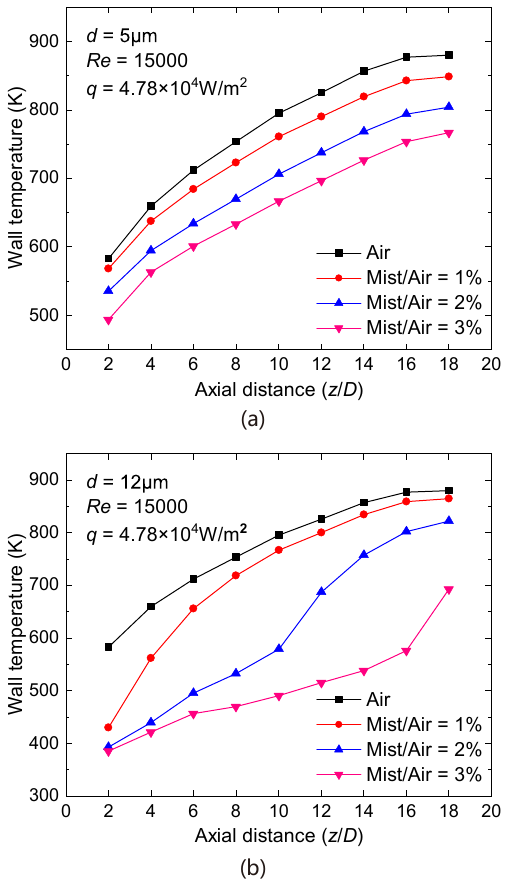}\\
  \caption{Wall temperature distribution of the test section at different mist/air mass ratios. (a) For the 5-$\upmu$m atomizer; (b) for the 12-$\upmu$m atomizer.}\label{fig:fig05}
\end{figure}

The wall temperature distribution along the heated test section at a typical condition is compared between air-only cooling and mist/air cooling with different mist/air mass ratios, as shown in Fig.\ \ref{fig:fig05}(a). With air-only cooling, the temperature increases gradually along the test section, and the range of the temperature variation is from a minimum temperature of 583 K at the inlet to a maximum temperature of 880 K at the outlet. By introducing the mist/air cooling, the temperature in the whole test section decreases, and the reduction in the temperature becomes greater if the mist/air mass ratio increases. At the mist/air mass ratio of 3\%, the reduction in the maximum temperature (at the outlet) is 113 K, and the reduction in the mean temperature along the channel is 115 K, indicating that the mist/air cooling is a very effective strategy in convective heat transfer. In addition, by introducing the mist/air at the mass ratio of 3\%, the temperature difference between the outlet and the inlet ${{T}_{w,out}}-{{T}_{w,in}}$ decreases from 297 K to 274 K, indicating that the temperature distribution under mist/air cooling is more uniform than air-only cooling.

The effect of the mist/air cooling using the 12-$\upmu$m atomizer is slightly different, even though the overall effect of the temperature reduction by the mist/air cooling is the same. The effect of the mist/air cooling is stronger when using the 12-$\upmu$m atomizer, particularly near the inlet region. By introducing the mist at 1\%, the temperature reduction at the inlet is 153 K, while the temperature reduction at the outlet is 15 K. If the mist ratio is increased to 2\%, we can see from the temperature distribution that heat transfer can be divided into two regions: the upstream region and the downstream region, each takes about half of the channel length. In the upstream region ($z<11D$), the temperature reduction is remarkable, with an average reduction of 213 K. In contrast, in the downstream region ($z>11D$), the temperature reduction is less significant, with an average reduction of 117 K. If further increasing the mist/air ratio to 3\%, the region with significant temperature reduction extends downstream and almost takes the entire channel.

\begin{figure}
  \centering
  \includegraphics[width=0.35\columnwidth]{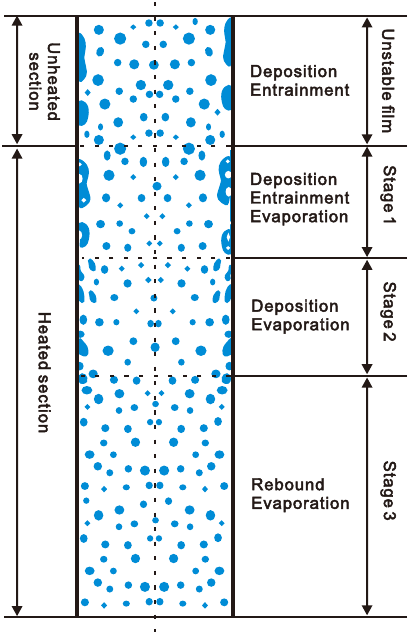}\\
  \caption{Schematic diagram of the mist/air cooling.}\label{fig:fig06}
\end{figure}

To explain the different effects of the mist/air flow on the heat transfer enhancement shown in Fig.\ \ref{fig:fig05}, the heat transfer mechanism of the mist/air two-phase flow is illustrated in Fig.\ \ref{fig:fig06}. In the experiment, the mist firstly passes through an unheated section before entering the heated test section. In the unheated section, some droplets inevitably deposit on the wall of the channel, forming larger droplets or even liquid films. When the larger droplets or the liquid films enter the heated test section by the entrainment effect of the air flow, they can induce a remarkable cooling effect because of the latent heat of the liquid. As it moves downstream, the entrainment effect reduces and the amount of large droplets and liquid films decreases. Hence, the deposition of droplets is mainly from the mainstream of the mist/air flow.

According to the interaction between the wall and the mist droplets, the cooling process of the mist/air flow in the heated section can be divided into three stages \cite{Guo2000ExperimentalSystem}. In each stage, the heat transfer equilibrium is affected by the wall heat flux, the droplet deposition and evaporation rate, and the airflow rate. In the first stage, the unstable liquid film and the coalesced droplets on the wall begin to boil. In this stage, droplets continue to be deposited on the wall, coalesce into large droplets, and flow downward in a boiling state under the entrainment of the airflow. In the second stage, as the wall temperature increases, the evaporation rate of the droplets increases. Even though droplets continue to deposit on the wall, the rapid evaporation of the droplets avoids the coalescence and the formation of large droplets. Without large droplets on the wall, the cooling effect of the mist decreases in comparison with the first stage. In the third stage, the wall temperature increases above the Leidenfrost temperature, then droplets will not be able to deposit on the wall, but bounce off from the wall quickly after the impingement. Even though the mist droplets do not deposit on the wall, as the air temperature is also high, the droplets evaporate gradually in the heated air, reducing the air temperature significantly by utilizing the latent heat of the droplets. Even though the heat transfer enhancement by the mist in the third stage is not as good as that in the first and the second stages, the mist can also improve the heat transfer efficiency due to the large latent heat of water droplets.

These three stages do not necessarily exist simultaneously in all experiments for the heated test section. When the droplet size is very small, the direct deposition of droplets on the wall decreases significantly and the temperature of the wall increases steeply at the entrance region. Hence, the first and the second stages will be short, and the mist/air two-phase flow in the test section is dominated by the third stage. For the heat transfer processes corresponding to Fig.\ \ref{fig:fig05}(a), the mist/air two-phase flow is in the third stage throughout the test section because the droplet size is very small. In this situation, the mist/air flow not only enhances the heat transfer efficiency of the test section but also improves the overall uniformity of the wall temperature distribution of the test section. In contrast, for the heat transfer processes corresponding to Fig.\ \ref{fig:fig05}(b), the mist/air flow contains larger droplets. The larger droplets are more likely to deposit on the wall of the unheated section and then enter the test section under the entrainment effect of airflow. It leads to a significant decrease in the temperature at the entrance region of the test section. The temperature reduction in this region leads to the deposition of more droplets on the wall, which further reduces the temperature of the wall. In contrast, after the entrance region, the wall temperature increases not only because the entrainment effect disappears, but also because the droplet deposition rate decreases after the depletion of large droplets. As a consequence, there is a steep rise in the wall temperature, as shown in Fig.\ \ref{fig:fig05}(b).

\begin{figure}
  \centering
  \includegraphics[width=0.65\columnwidth]{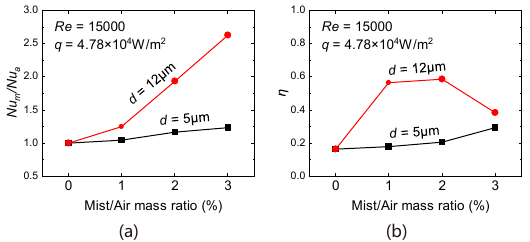}\\
  \caption{(a) Thermal performance ratio ${{{\Nu}_{m}}}/{{{\Nu}_{a}}}$ and (b) heat transfer non-uniformity $\eta$ of the test section varying with the mist/air mass ratio for different droplet sizes.}\label{fig:fig07}
\end{figure}

The performance of the mist/air cooling can be compared quantitatively using the thermal performance ratio and the heat transfer non-uniformity, as shown in Fig.\ \ref{fig:fig07}. The thermal performance ratio increases as the mist/air ratio increases, as shown in Fig.\ \ref{fig:fig07}(a). As the mist/air mass ratio increases from 0 (air-only cooling) to 3\%, the thermal performance ratio increases to 1.23 for the 5-$\upmu$m atomizer, and increases to 2.62 for the 12-$\upmu$m atomizer. This result indicates that the mist/air cooling is very effective, particularly for larger droplets. Mist/air cooling has a higher heat transfer non-uniformity compared to air-only cooling, as shown in Fig.\ \ref{fig:fig07}(b). As the mist/air mass ratio increases, the heat transfer non-uniformity for the 5-$\upmu$m atomizer gradually increases. For the 12-$\upmu$m atomizer, however, the heat transfer non-uniformity increases significantly when introducing the mist/air cooling at the mass ratio 1\%. This is mainly due to the significant heat transfer enhancement near the entrance, which leads to the remarkable temperature reduction near the entrance and less temperature reduction in the downstream. When the mist/air mass ratio increases to 2\%, the heat transfer non-uniformity increases slightly as the temperature reduction region propagates towards the downstream. When the mist/air mass ratio increases to 3\%, the heat transfer non-uniformity significantly decreases, because the region with significant temperature reduction extends downstream and almost takes the entire channel.

\subsection{Effect of wall heat flux and heat transfer capability}\label{sec:sec32}
\begin{figure}
  \centering
  \includegraphics[width=0.5\columnwidth]{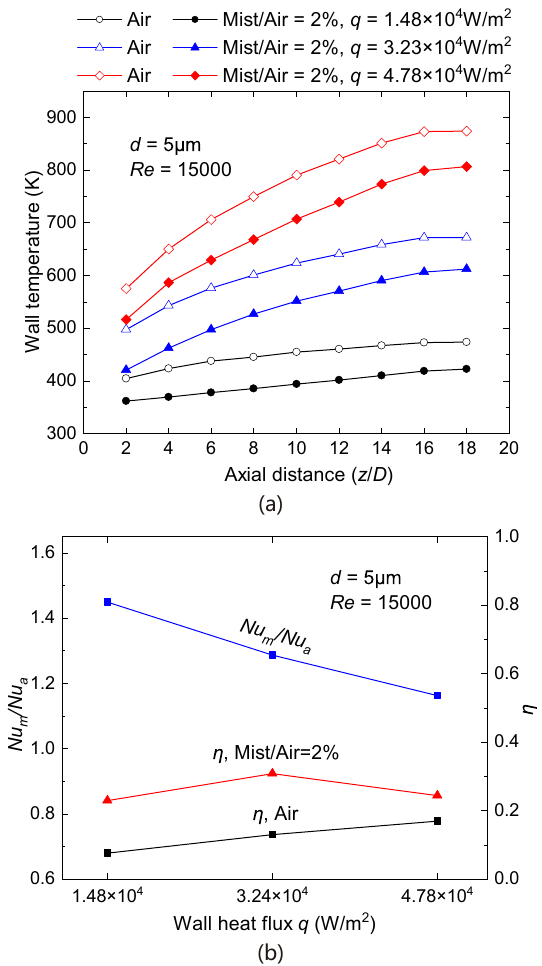}\\
  \caption{Effect of wall heat flux on the mist/air cooling. (a) Wall temperature distribution with different wall heat fluxes; (b) Thermal performance ratio and heat transfer non-uniformity for different wall heat fluxes.}\label{fig:fig08}
\end{figure}

The wall temperature distributions of the heated test section at different wall heat fluxes are compared between mist/air flow and air-only flow, as shown in Fig.\ \ref{fig:fig08}. The variation of the wall temperature along the channel is relatively small at low wall heat flux, but the variation becomes steep as the wall heat flux increases. Compared with that in air-only cooling, the wall temperature in air/mist cooling decreases significantly for all the wall heat fluxes considered in the study. As the heat flux increases, the temperature reduction by the mist/air cooling increases. When the wall heat flux increases from $1.48\times10^4$ to $4.78\times10^4$ W/m$^2$, the reduction of the mean temperature by the mist/air cooling increases from 55 to 74 K, and the reduction of the maximum temperature by the mist/air cooling increases from 50 to 67 K. These results indicate that the mist/air flow is an effective cooling strategy particularly for high-heat-flux cooling applications.

To quantitatively consider the effect of the heat flux on the heat transfer performance, the thermal performance ratio and the heat transfer non-uniformity are plotted in Fig.\ \ref{fig:fig08}(b). As the wall heat flux increases, the thermal performance ratio gradually decreases. This is because as the wall heat flux increases, the wall temperature increases and the droplet-wall interaction (as illustrated in Fig.\ \ref{fig:fig06}) changes from the first stage to the third stage. Although the increase in air temperature can accelerate the droplet evaporation, the direct contact between the droplets and the wall has a greater influence on the heat transfer of the test section. As a result, with the increase of the wall heat flux, the heat transfer enhancement is gradually weakened.

As the wall heat flux increases, the heat transfer non-uniformity increases continuously for air-only cooling and varies slightly for mist/air cooling. For air-only flow, as the heat flux increases, the wall temperature increases and the cooling performance in the downstream decreases. Hence, the heat transfer non-uniformity increases. In contrast, for mist/air flow, the overall heat transfer non-uniformity is higher than that for air-only flow. But when the heat flux is very high, almost the entire channel is in the third stage, which reduces the direct deposition of droplets on the wall and reduces the heat transfer non-uniformity.

\begin{figure}
  \centering
  \includegraphics[width=0.5\columnwidth]{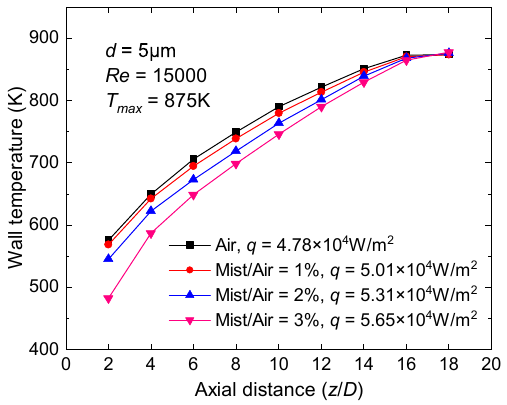}\\
  \caption{Wall temperature distribution along the test section for the test of the heat transfer capability at different mist/air mass ratios. The heat flux was varied to maintain the highest wall temperature (at the outlet) fixed at 875K. }\label{fig:fig09}
\end{figure}

In many applications, it is important to maintain the wall temperature below a threshold, beyond which the system may not function properly or the material may fail. Considering this concept, experiments were performed by increasing the heat flux until the maximum temperature (at the outlet) reaches a predetermined value. In such a way, the capability of the mist/air cooling at different mist/air ratios is tested and compared with that of air-only flow, as shown in Fig.\ \ref{fig:fig09}. The results show that, as compared with the air-only cooling, the wall heat flux of the mist/air cooling with the mist/air mass ratio of 1\%, 2\%, and 3\% increases by 4.8\%, 11.1\%, and 18.4\%, respectively. Therefore, a small amount of mist in air can increase the heat transfer capacity significantly.

\subsection{Effect of Reynolds number}\label{sec:sec33}
\begin{figure}
  \centering
  \includegraphics[width=0.5\columnwidth]{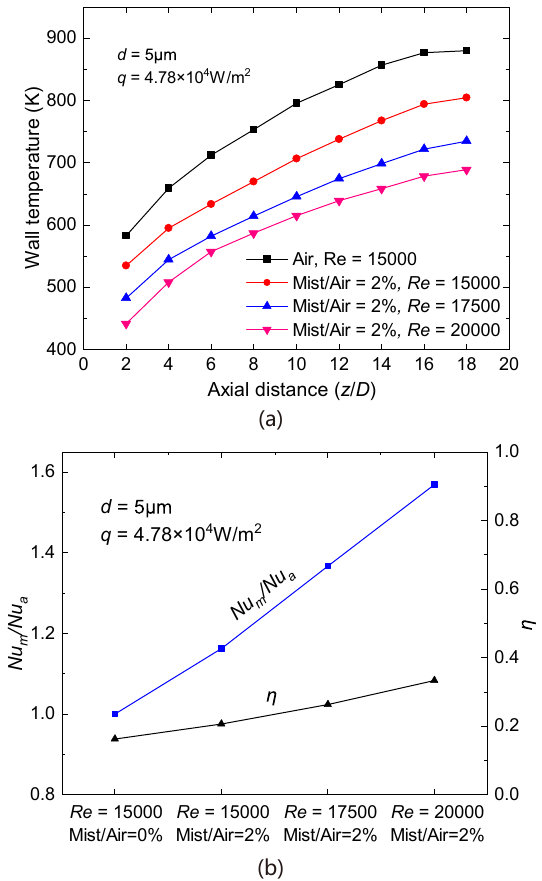}\\
  \caption{Effect of Reynolds number on the mist/air cooling. (a) Wall temperature distribution variation; (b) Thermal performance ratio and heat transfer non-uniformity of the test section.}\label{fig:fig10}
\end{figure}

The temperature distribution of the heated test section at different Reynolds numbers is shown in Fig.\ \ref{fig:fig10}(a). As the Reynolds number increases, the overall wall temperature of the test section decreases, but the temperature reduction is more significant in the downstream than that in the upstream. The reason is that the wall temperature at the entrance region is lower and the temperature difference with the coolant is smaller than that in the downstream (because of the effective heat transfer by the higher mist concentration and the droplet-wall interaction). Therefore, when the Reynolds number increases, the potential of the temperature reduction at the entrance region is limited as the temperature difference is already very small. In contrast, the wall temperature at the exit region of the test section is much higher, allowing more space for temperature reduction if the heat transfer is enhanced. As the Reynolds number increases, the temperature difference between the wall and the coolant decreases, i.e., a more significant reduction in the wall temperature at the exit region. With the reduced temperature difference between the wall and the coolant at high Reynolds numbers, the effect of further increasing the Reynolds number is limited. Therefore, as the Reynolds number increases, the effectiveness of increasing $Re$ on reducing the wall temperature (i.e., $-dT/dRe$) becomes smaller. Therefore, considering the extra flow resistance produced by the higher Reynolds number, the optimum condition may not be at the highest Reynolds number since the benefit of increasing the Reynolds number for heat transfer enhancement decreases as the Reynolds number increases. A moderate Reynolds number may result in a satisfactory temperature reduction with reasonable flow resistance.

With increasing the Reynolds number, both the thermal performance ratio and the heat transfer non-uniformity increase, as shown in Fig.\ \ref{fig:fig10}(b). With increasing the Reynolds number, both the air mass flow rate and the mist mass flow rate increase, so the heat transfer capacity of the mist/air flow increases significantly and the wall temperature decreases significantly. With the increase of the mass flow rate of mist and the decrease of the wall temperature, the direct deposition of droplets on the wall increases, resulting in heat transfer enhancement. Since the deposition of the droplets on the wall is mainly in the entrance region, the heat transfer non-uniformity also increases.

\section{Conclusions}\label{sec:sec4}
The heat transfer of mist/air cooling in a high-temperature channel is studied experimentally in this work. New atomizers were designed to facilitate the atomization of water into droplets of different sizes with accurate control of the mist/air mass ratio. A series of experiments were carried out to investigate the effects of the mist/air mass ratios, the Reynolds number, the wall heat flux, and the droplet sizes on the heat transfer performance.

The results of the study show that mist/air cooling is an effective strategy for heat transfer enhancement in high-temperature channels. Only a small amount of droplets in mist/air cooling can significantly enhance the cooling performance. With large droplets, the temperature reduction is more remarkable in the upstream than in the downstream. This feature is because the cooling process of the mist/air flow in the heated section can be divided into three stages according to the interaction between the wall and the mist droplets. With increasing the mist/air ratio, the region with significant temperature reduction extends downstream. For mist droplets with different sizes, larger droplets can induce more significant temperature reduction, while smaller droplets can improve the temperature distribution uniformity of the test section. The capability of the mist/air cooling was tested by increasing the heat flux until the maximum temperature (at the outlet of the test section) reaches a predetermined value, and as compared with the air-only cooling, the increment in the wall heat flux by the mist/air cooling with a mass ratio of 3\% can be up to 18.4\%. As the Reynolds number increases, the reduction rate of the overall temperature of the test section becomes smaller, and the benefit of increasing the Reynolds number for the heat transfer enhancement decreases. Hence, a moderate Reynolds number may result in a satisfactory temperature reduction with reasonable flow resistance.

The mist/air flow is a complex process involving many spatial and temporal scales, and deserves further detailed studies, such as the direct visualization of the mist/air flow in the channel, the quantitative measurement of the interaction between the droplet and the channel, and the across-scale simulations of the flow and the heat transfer. The understanding of the process will be useful for the optimization of the cooling process of high-temperature surfaces in relevant applications.

\section*{Declaration of Competing Interest}
None.

\section*{Acknowledgements}
This work was supported by the Key Laboratory Fund (6142702200509) and the National Natural Science Foundation of China (Grant No.\ 51676137).


\bibliography{MistAirCooling}

\end{document}